\documentclass{hep99}

\usepackage[dvips]{graphicx}
\usepackage{epsf}
\begin{document}

\twocolumn[
\begin{center}
\centerline{\LARGE\bf
Charm hadroproduction results from Selex
}
\vskip 0.3cm
{\Large 
\centerline{The SELEX Collaboration}
M.~Iori$^{r}$,
{\small N.~Akchurin$^{p}$,
V.A.~Andreev$^{k}$,
A.G.~Atamantchouk$^{k}$,
M.~Aykac$^{p}$,
M.Y.~Balatz$^{h}$,
N.F.~Bondar$^{k}$,
A.~Bravar$^{t}$,
P.S.~Cooper$^{e}$,
L.J.~Dauwe$^{q}$
% ...
\\ 
% ...
$^e$Fermilab, Batavia, IL 60510, U.S.A.\\
% ...
$^h$Institute of Theoretical and Experimental Physics, Moscow, Russia\\
% ...
$^k$Petersburg Nuclear Physics Institute, St. Petersburg, Russia\\
% ...
$^p$University of Iowa, Iowa City, IA 52242, U.S.A.\\
$^q$University of Michigan-Flint, Flint, MI 48502, U.S.A.\\
$^r$University of Rome ``La Sapienza'' and INFN, Rome, Italy\\
% ...
$^t$University of Trieste and INFN, Trieste, Italy\\
}
}
\end{center}
\vskip 0.3cm

\centerline{\bf Abstract}{
The SELEX experiment (E781)  is 3-stage magnetic spectrometer for
   a high statistics study of hadroproduction of charm baryons out to
   large $x_{F}$ using 650 Gev $\Sigma ^{-}$, $\pi ^{-}$ and $p$ beams. 
    The main features
   of the spectrometer are: a high precision silicon vertex system, 
    powerful particle identification provided by TRD
   and RICH, forward $\Lambda _{s}$ decay spectrometer and 3-stage 
   lead glass photon
   detector.
   An experiment overview  and spectrometer features
   are shown. Reconstructed charm states  and
   results on $\Lambda _{c}$, $D^{+}$ particles and antiparticles
   produced by $\Sigma ^{-}$, $\pi ^{-}$ and $p$ beams at $x_{F}>0.3$ 
   and asymmetry for $\Lambda _{c}$ are presented.
} 

\vskip 0.3cm

]


\section{Introduction}

\begin{figure}
\epsfxsize=\hsize
\epsffile{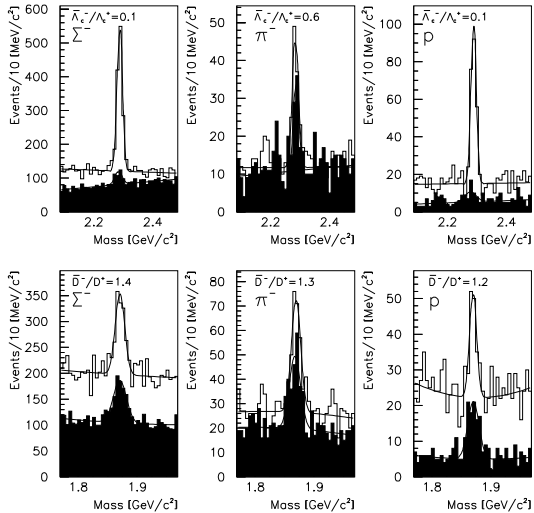}

\caption{Charm baryon and meson asymmetry integrated over $x _{F} >0.3$
for different beams}

\end{figure}

Charm physics explores QCD phenomenology in both perturbative 
and nonperturbative
regimes. Production dynamics studies test leading order (LO)
and next to leading order (NLO) perturbative QCD.
Charm lifetime measurements test models based on $1/M_{Q}$ QCD 
expansions. The present fixed target experiments have considerably
improved the statistics but many problems remain. 
All experimental results are in qualitative agreement with perturbative QCD
calculations but quantitative deviations from QCD are observed~\cite{Frix97}.
More experimental data, using different incident hadrons ($\pi$, $p$ and
$\Sigma ^{-}$), help to discriminate possible different scenarios:
dragging effects, fragmentation effects as well as intrinsic 
$k _{t}$~\cite{theo}.

\section{The Selex spectrometer}
The SELEX experiment at Fermilab is a 3-stage magnetic spectrometer.
The 600 Gev/$c$ Hyperon beam of negative polarity 
contains equal fraction of 
$\Sigma$ and $\pi$.  The positive 
beam is composed 92\% of protons and the rest $\pi$'s.
Beam particles are identified by a Transition Radiation detector (BTRD).
The spectrometer was designed to study charm production in the forward 
hemisphere with good mass and decay vertex resolution for charm momentum
in a range of 100-500 Gev/$c$.
The vertex region is composed of 5 targets (2 Cu and 3 C ). The total
target thickess is 5\% of $\lambda _{int}$ for protons and the targets are
separated by 1.5 cm. Downstream of the targets there are 20  
silicon planes with a strip pitch of 20-25 $\mu$m disposed in X,Y,U and
V views.
The M1 and M2 magnets effect a momentum cut off of 2.5 Gev/$c$ and 15 Gev/$c$ 
respectively. A RICH detector, filled with Neon at room temperature and pressure, 
provides a single track ring radius resolution of 1.4\% and
 2$\sigma$ $K/ \pi$ separation up to about
185 Gev/$c$. A computational filter uses   tracks identified by the RICH 
and linked to the vertex silicon by the PWCs to make a full reconstruction
of the secondary vertex. Events consistent with only a primary vertex
are rejected.
A Transition Radiation Detector (TRD) provides electron identification. 

\section{Data set and charm selection}

The charm trigger is
very loose. It requires a valid beam track, two tracks of opposite charge 
to the beam with momentum $>15$ Gev/$c$, 
two high momentum
and linked to the Silicon vertex detector, at least one of which is
displaced from the primary vertex.
We triggered on about 1/3 of all
the inelastic interactions. About 1/8  of them are written on the tape
for a final sample of about 1.0B events.
In the analysis secondary vertices were reconstructed if the
$\chi ^{2}$ of all tracks was inconsistent with single primary vertex.
The RICH detector labelled all particles above 25 Gev/$c$.

All data reported here resulted from a preliminary pass through
the data, using a production code optimized for speed and not for
efficiency. The simulated reconstruction efficiency of any charmed state
is  constant at about 40\% for $x_{F}>0.3$.

\begin{figure}
\vspace{5.7cm}
\includegraphics{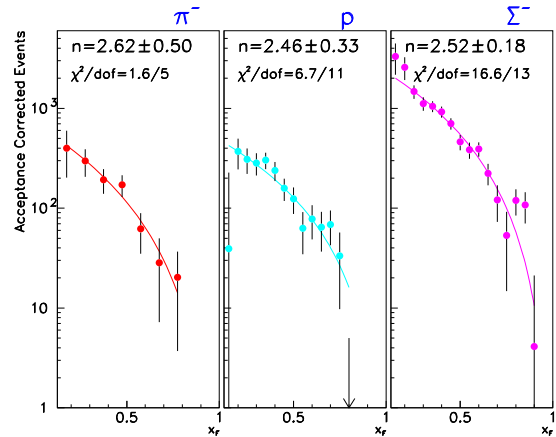}
\caption{  ${\Lambda _{c}}^{+}$ $x _{F}$ dependence for different beams}
\end{figure}

\begin{figure}
\vspace{6cm}
\includegraphics{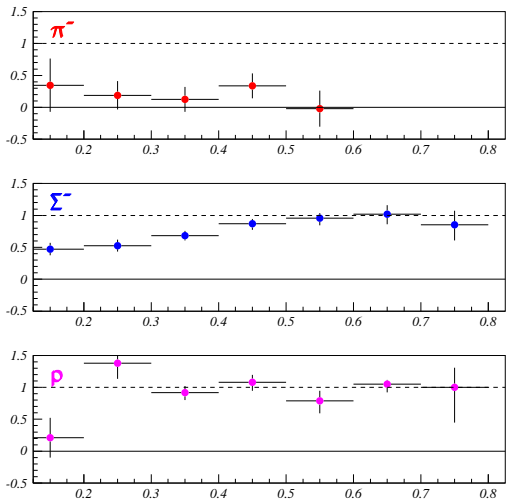}
\caption{  ${\Lambda _{c}}^{+}$ asymmetry versus $x _{F}$ for different beams.
The asymmetry is $A={{N_{\Lambda _{c}}} - {N_{\bar \Lambda
 _{c}}}}/{{N_{\Lambda _{c}}} + {N_{\bar \Lambda _{c}}}}$}
\end{figure}
\subsection{Charm performance}

The requirements to study charm physics and to reduce the
background are: 
\par
good decay vertex resolution and mass 
resolution and particle identification. 
\par
The vertex algorithm
provides an average longitudinal error, $\sigma _{z}$ on the primary
and secondary vertex of 270 $\mu$m and 500$\mu$m respectively. Their
combination, $\sigma$ is equal to 570 $\mu$m. In the  $\Lambda _{c}$
sample, the average momentum is 220 Gev/$c$, 
corresponding to average decay length of 7.5 $mm$.
\par
To reduce the background we only keep charm candidates that decay
over a longitudinal distance greater than at least $z _{min}= 8 \sigma $
\par
The RICH detector gives a good
separation $\pi /K$ for track momentum of 100 Gev/$c$ corresponding
to the average momentum of decay tracks (p, K).

The mass resolution is constant versus the momentum (i.e the 
$D ^{0} \rightarrow K ^{-} + \pi ^{+}$ mass resolution is $<$ 10 Mev/c
at all momenta).
\section{Charm and anticharm asymmetry}

To lowest order QCD, charm and anticharm quarks are 
produced symmetrically in hadroproduction. Next to Leading Order
(NLO) introduces small asymmetries in quark momenta due to
interference between contributing amplidutes. 
In this connection
the ACCMOR experiment does not find distinction between leading and
nonleading charm meson and charm baryons produced in a 200 Gev/$c$ pion beam and 
a 250 Gev/$c$ proton beam~\cite{ACCM}.
\par
However E769  has observed a $\Lambda _{c}$ asymmetry integrated 
over $x _{F} >0$
for a 250 Gev/$c$ proton beam and in the same experiment has measured
D meson asymmetry for a 250 Gev/$c$ pion beam~\cite{E769}.
\par
The WA89 experiment has studied charm particles produced by a $\Sigma ^{-}$
340 Gev/$c$ beam. Considerable production asymmetry between $D ^{-}$,
$D ^{+}$ and ${\Lambda_{c} }^{+} $, ${\overline{\Lambda_{c}}}^{-}$ 
was observed~\cite{WA89}.
\par
Recently the WA92 and E791 experiment show charm D meson asymmetry in a 350 Gev/$c$
pion beam~\cite{WA92} and a 500 Gev/$c$ pion beam~\cite{E791} respectively.
\par
The observed asymmetry can be explained by a recombination of charm
quark antiquarks with the beam valence quarks or by different
processes like in the intrinsic charm and in the quark-gluon 
string model~\cite{theo}.
\par
In the SELEX experiment the negative beam particles have a
valence quark in common with ${\Lambda _{c}}^{+}, D^{-}, {D _{s}}^{-}$ 
(d or s quark) or with $ D^{0}$ (u quark). 
\par
The 
proton beam has two valence quarks in common with  ${\Lambda _{c}}^{+}$
(u and d quarks) and one quark with $D^{-}$ (d quark).
The $\pi$ data show comparable particle-antiparticle yields both
for charm mesons and charm baryons. 
The observed $D^{\pm}$ meson asymmetry at $x _{F}>0.4$ 
does not rise steeply with $x _{F}$ as previously reported~\cite{WA92,E791}.
\par
For both baryon beams  ($\Sigma ^{-}, p$) there is a difference in
leading-nonleading production integrated over $x _{F}>0.3$, of charm 
mesons and charm baryons.
In particular they show asymmetric particle-antiparticle yields for
 ${\Lambda _{c}}^{+}$ and almost symmetric yields for $D^{\pm}$ mesons.
The results shown in Fig.~1 are consistent with previous low statistics 
measurements~\cite{E769,WA89}.  

Fig.~2 shows the $\Lambda _{c} ^{+}$ $x _{F}$ dependence for different beams.
There is a clear evidence that $\Lambda _{c} ^{+}$ is leading particle 
for all three beams and the n-dependence is quite hard and essentially
the same for the three incident hadrons.
Fig.~3 shows the $\Lambda _{c} ^{+}$ asymmetry versus $x _{F}$. 
The asymmetry is clearly large for the baryon beams than for the $\pi$
beam. For the protons the only region in which there is $\bar \Lambda _{c}$
production is at very small $x _{F}$.

\section{SUMMARY}
The SELEX experiment explores the large $x _{F}$ region using different beams. 
The present results show new features in charm production that 
complement the experiments using a lower energy beam (200-350 Gev/$c$).
Clear evidence for leading $\Lambda _{c}$ was observed. Other
charm baryon states are being analyzed and will be reported later.

\end{document}